\documentstyle[aps,floats,pre]{revtex}
\input{epsf}

\begin{document}
\bibliographystyle{h-physrev}

\newcommand{\qmodel}{$q$-model}
\newcommand{\qMODEL}{$q$-MODEL}
\newcommand{\by}{\times}
\newcommand{\sright}{s_{ij}^{+}}
\newcommand{\sleft}{s_{ij}^{-}}
\newcommand{\sboth}{s_{ij}^{\pm}}
\newcommand{\vboth}{v_{ij}^{\pm}}
\newcommand{\fright}{f_{ij}^{+}}
\newcommand{\fleft}{f_{ij}^{-}}
\newcommand{\fboth}{f_{ij}^{\pm}}
\newcommand{\Iboth}{I_{ij}^{\pm}}

\title{Force distribution in a scalar model for non-cohesive granular material}
\author{Matthew~G.~Sexton$^1$, 
        Joshua~E.~S.~Socolar$^{1,3}$
        David~G.~Schaeffer$^{2,3}$} 
\address{$^1$Department of Physics, $^2$Department of Mathematics, and
         $^3$Center for Nonlinear and Complex Systems,\\
         Duke University, Durham, NC 27708} 
\date{\today}

\maketitle

\begin{abstract}
  We study a scalar lattice model for inter-grain forces in static,
  non-cohesive, granular materials, obtaining two primary results.
  (i) The applied stress as a function of overall strain shows a power
  law dependence with a nontrivial exponent, which moreover varies
  with system geometry.  (ii) Probability distributions for forces on
  individual grains appear Gaussian at all stages of compression,
  showing no evidence of exponential tails.  With regard to both
  results, we identify correlations responsible for deviations from
  previously suggested theories.
\end{abstract}


\section{Introduction}
Attempts to understand stress distribution in static,
non-cohesive, granular materials have uncovered a rich structure of
force chains that currently is poorly understood \cite{nagel}.
Although mechanical systems such as bead or granular packings 
subject to uniaxial compression seem straightforward, 
they harbor fundamental theoretical challenges concerning 
the connection between the micro and macro scales.  
Such systems are far from thermal equilibrium since
thermal energy scales are negligible compared to the potential
energies stored in elastic deformation of the grains.  In addition,
the force between grains is strongly nonlinear, as it vanishes
identically when the grains are not in contact.

\begin{figure}[htbp]
  \centerline{ \epsfxsize=4.0in \epsfbox[0 0 364 372]{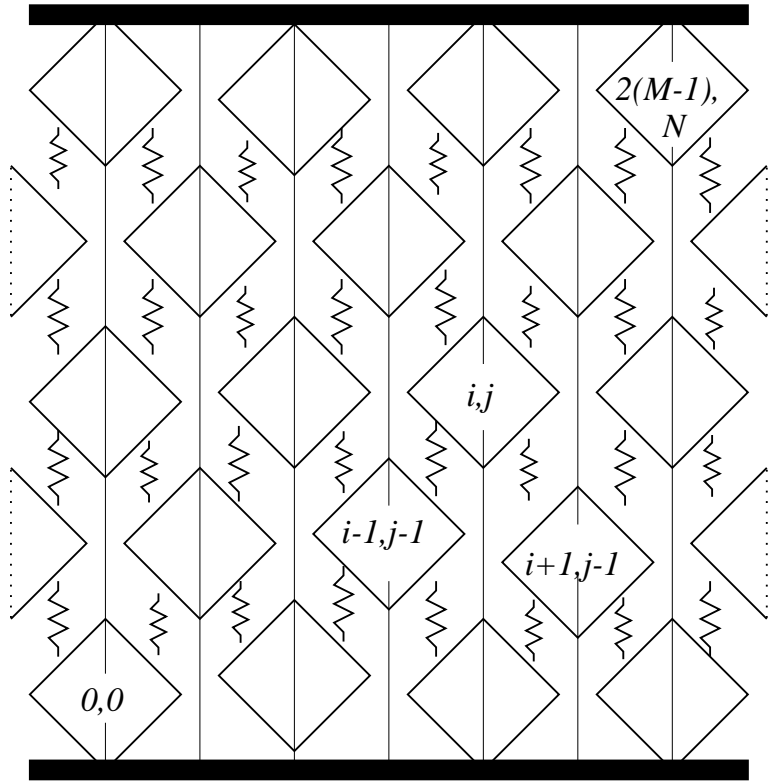} }
  \vspace{0.2in} \caption{The scalar model.  Diamonds represent rigid
  grains that slide on frictionless, vertical wires.  Squiggly lines
  represent springs with identical spring constants but differing
  equilibrium lengths.  Grains cannot penetrate the ceiling or floor.}
  \label{fig:model}
\end{figure}

In this paper, we report results and analyses of a toy model for the
compaction of a granular material.  Our model consists of ``grains''
placed on a square lattice and connected by vertical springs.  The
grains are constrained to move only in the vertical direction, so the
model treats forces and stresses as {\em scalar} quantities.  (See
Fig.~\ref{fig:model}.)  Geometric disorder is introduced in the model
by assigning to each spring an equilibrium length drawn randomly from
a square distribution.  Though we cannot hope to capture the true
physics of tensorial stresses in this manner, we can study the
validity of arguments that have been applied to real granular systems
and should equally well apply to our toy model.  In particular, we
compare the distribution of inter-grain forces generated by our model
to the distribution predicted by the
\qmodel\ of Coppersmith et al \cite{qmodel1},
and we compare the stress-strain power law generated by
our model to that expected on the basis of a mean-field argument.
Our results provide a warning:
in neither case are the theoretical predictions borne out by the numerics.

\section{Definition of the model}\label{sec:modeldef}
The structure of our model is shown in Fig.~\ref{fig:model}.  
Grains are placed at the even-parity sites of a $2 N_x\by (N_y+1)$ cubic
lattice with lattice constant $b$.  We refer to such a system as an
``$N_x\by N_y$'' system.  Excluding the sites on the lower boundary,
it contains $N_x N_y$ grains.  Grains are constrained to move only
in the vertical direction, as if sliding on fixed,
frictionless wires (shown as thin lines in the figure).  We let
$u_{i,j}$ denote the vertical displacement of the grain at lattice
site $(i,j)$ from its nominal height $jb$, 
with $u_{i,j}>0$ for upward displacements.  
Between every two nearest-neighbor 
grains there is a vertical spring.  All of the springs
respond linearly under compression, with spring constant $k$, but they
produce no force under extension.  The energy of the spring between
grains $i,j$ and $i\pm 1,j-1$ is taken to be
\begin{equation}\label{eq:Eij}
  E_{ij}^{\pm} = \left\{ \begin{array}{ll}
                 \frac{1}{2}k\left(u_{i,j} -\sboth -u_{i\pm 1,j-1}\right)^2 
               &  \mbox{if} \hspace{1em}
                        (u_{i,j} -\sboth -u_{i\pm 1,j-1}) < 0 \\
           0  &  \mbox{otherwise}
           \end{array}\right. 
\end{equation}
where $\sright$ and $\sleft$ are independent, quenched random
variables designating the different uncompressed lengths of the
springs.  We take $\sright$ and $\sleft$ to be distributed uniformly
over the interval $(-\epsilon,\epsilon)$.  Note that when $s=0$,
the uncompressed spring has length $b$, the lattice constant.
One may imagine the grains to have arbitrary shapes and the
uncompressed spring length to be the vertical spacing between grain
centers when the grains are just barely in contact.

The system is taken to satisfy periodic boundary conditions in the
horizontal direction.  
All grains at the top are constrained from above by a rigid ``ceiling''
and therefore satisfy $u_{i,N_y}\leq 0$.
On the bottom row, grains are constrained by a rigid ``floor'' whose height $U$
is an independent variable; thus $u_{i,0} \ge U$.
Let $U_0$ denote the maximum value of $U$ such that no springs are
compressed.
Typically $U_0$, which is a function of the random $\sboth$'s, 
is negative and of the order of $N_y\epsilon$ in absolute value.
(It cannot be less than $-N_y\epsilon$.)
Below we measure the floor height in terms of the incremental variable
$\Delta\equiv U-U_0$.

Though we use the terms ``floor'', ``vertical'', etc., for
convenience, there is no gravity in the model and there is complete
statistical symmetry between the up and down directions.  The
algorithm discussed below for determining equilibrium configurations
explicitly breaks this symmetry, though the configurations it produces
do not.  

\subsection{Meaningful parameters}
Our model apparently depends on the parameters $\epsilon$ and $k$,
but these merely set the scales for distance and force.
For the rest of this paper, numerical values for distances ---
spring-length variations $\sboth$, grain displacements $u_{i,j}$,
and boundary displacement $\Delta$ ---
will be quoted in units of $\epsilon$; 
and forces ---
individual forces $\fboth$ and the total force $F$ on the ceiling ---
in units of $k\epsilon$.

Although the {\em width} of the distribution of spring lengths
scales out of the problem, its {\em shape} could conceivably make a difference.
Nevertheless, we do not expect the precise shape to be important as long as
the probability of having very large uncompressed lengths decays sufficiently
rapidly.  In any event, we consider only the square distribution.

Let us contrast the alternative of random spring constants with 
the random equilibrium lengths we study here.
If the $k$'s were random but all equilibrium lengths were equal,
then as $\Delta$ increased all the springs would begin to be compressed
simultaneously at $U=0$ and the system would be perfectly linear for
all $\Delta>0$;
the force-vs.-displacement law would be 
$F(\Delta)=K\Delta$ for $\Delta>0$ and $F(\Delta)=0$ for $\Delta<0$,
where $K$ is a constant determined by the random $k$'s.
By contrast, with random spring lengths 
we capture some of the geometric disorder of real granular materials,
which inevitably introduces nonlinearities associated with 
the formation of new contacts during compression.  

\subsection{General features of equilibrium configurations}
\label{subsec:features}
We will refer to a spring under compression as an ``active bond'', and
to a site connected to such a spring as an ``active site.''  
An equilibrium configuration for $\Delta$ positive, 
but not too large, 
has a mixture of active and inactive sites and bonds.  
For a given set of $\sboth$'s, 
the equilibrium configuration of active sites and bonds is unique,
although the exact positions of inactive sites are not determined. 
This uniqueness follows from the strict convexity of the total energy
as a function of the lengths of active bonds.

\begin{figure}[htbp]
  \centerline{ \epsfxsize=6.0in \epsfbox[0 0 444 147]{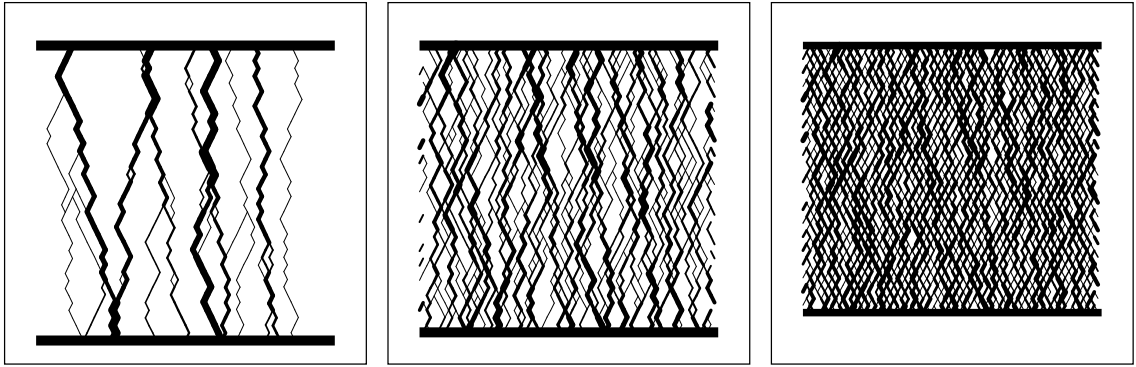}} 
  \vspace{0.2in} \caption{Configurations of active bonds in a 
  $40\by 40$ system at three different stages of compression.}
  \label{fig:configs}
\end{figure}

Fig.~\ref{fig:configs} shows three equilibrium configurations with the
same set of $\sboth$'s on a $40\by 40$ lattice, for
different amounts of compression.  Only the active bonds are
shown, with line thickness proportional to the force transmitted.
Also shown are two bars representing the ceiling and the floor.

\begin{figure}[htbp]
  \centerline{ \epsfxsize=6.0in \epsfbox[69 349 516 669]{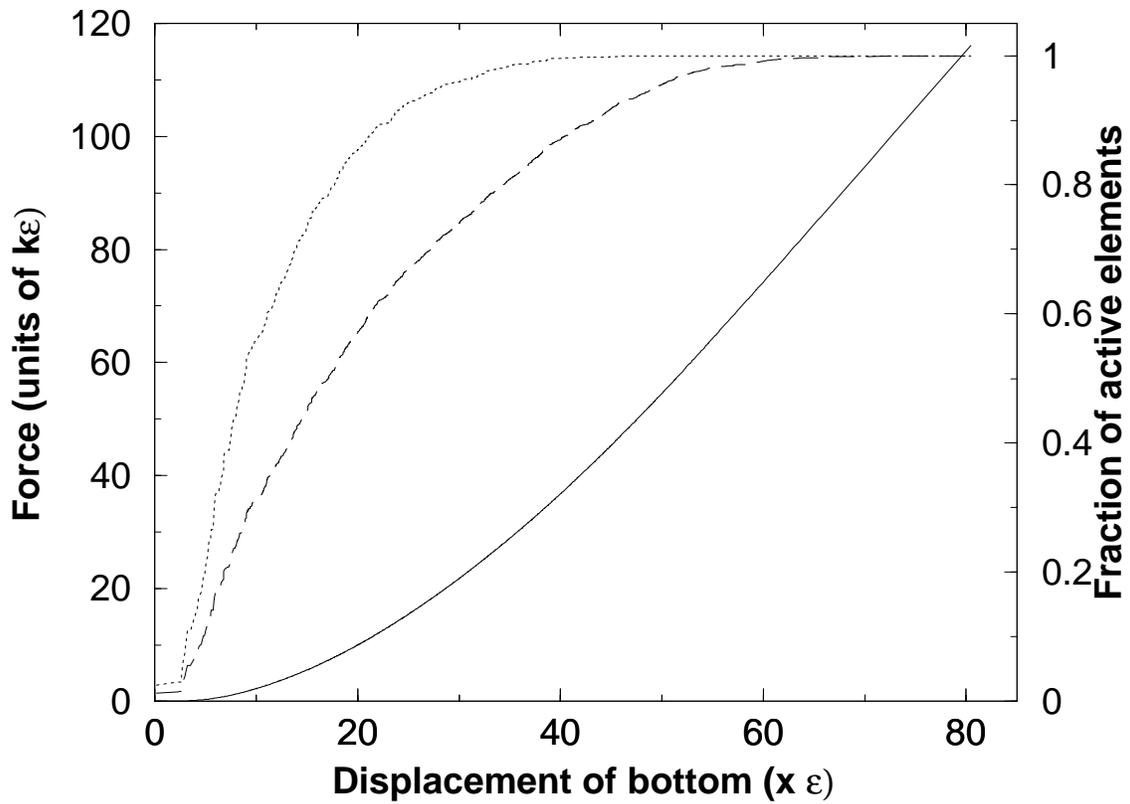}}
  \vspace{0.2in} \caption{A typical force curve (solid line) for a $40\by
  40$ system together with curves showing the fraction of active sites
  (dotted) and the fraction of active bonds (dashed). }
  \label{fig:onecurve}
\end{figure}

Let $F(\Delta)$ be the total force exerted by the system on the
ceiling.  Fig.~\ref{fig:onecurve} shows a typical force curve for a
$40\by 40$ system, together with the fractions of sites and bonds that
are activated.  Elementary considerations reveal the limiting behaviors of
$F$ for large and small $\Delta$.  For very small $\Delta$, the force
on the ceiling is due to a single chain of activated springs that
reaches from top to bottom.  
This chain, which follows the directed path
$i(j)$ through the lattice that yields the highest value of
$\sum_{j=1}^{N_y} s_{ij}$, produces a force linear in $\Delta$:
\begin{equation}\label{eq:FsmallD}
 F(\Delta) = \frac{1}{N_y}\Delta \quad \mbox{($\Delta$ small)}. 
\end{equation}
When $\Delta$ is very large, all of the springs in the system are
compressed.  The system is then again entirely linear:  
because all of the springs have the same stiffness, 
\begin{equation}\label{eq:FlargeD}
 F(\Delta) = \frac{2N_x}{N_y}\Delta + C \quad \mbox{($\Delta$ large)},
\end{equation}
where $C$ is a (negative) constant that depends on the
random $\sboth$'s and is difficult to compute.

\begin{figure}[htbp]
  \centerline{ \epsfxsize=3.5in \epsfbox[0 0 542 209]{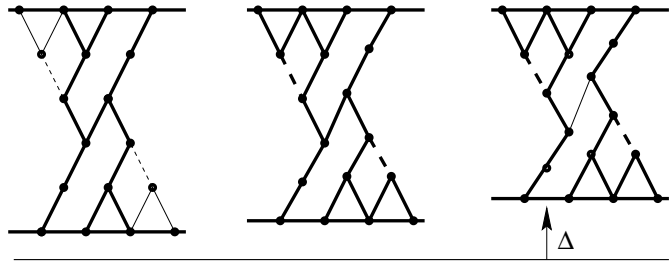} }
  \vspace{0.2in} \caption{A sequence of configurations in which the
  central bond is originally active, but becomes inactive after other
  bonds become active.  Consequently the total stiffness of the
  network decreases as the system is compressed.  Thick (thin) solid
  lines represent active (inactive) springs with $s=0$.  Thick (thin)
  dashed lines represent active (inactive) springs with equilibrium
  length $s=-\epsilon$.  Springs not drawn are assumed to have small
  enough equilibrium lengths that they are never active in this
  sequence.  $dF/d\Delta$ is smaller for the third configuration than
  for the second, although $\Delta$ itself has increased.}
  \label{fig:break}
\end{figure}

In general, as $\Delta$ is increased from zero, additional chains are
activated and $F(\Delta)$ increases.  In fact it can be shown that $F$
must increase monotonically with $\Delta$.  (This is why $C$
must be negative.)  Surprisingly, however, the slope of $F(\Delta)$
may {\em decrease} with increasing $\Delta$.  Occasionally bonds that
are active at small values of $\Delta$ become inactive for larger
$\Delta$, thus reducing the effective spring constant of the network.
The simplest example of this behavior is illustrated in
Fig.~\ref{fig:break}.  The figure shows a sequence of configurations
in which the central bond is originally active, but becomes inactive
due to the effective stiffening of other parts of the network when
additional bonds are activated.

\section{The numerical algorithm}
The algorithm we employ for generating equilibrium configurations of
our model relies heavily on the fact that for a
given set of active bonds, the system response is linear.  
By solving for the times at
which inactive bonds are activated during compression, the algorithm
generates the entire curve $F(\Delta)$, starting from $\Delta=0$ and
explicitly visiting every configuration.
To construct the initial, uncompressed configuration,
we begin with $u_{i,0}=U_0$ for all $i$ and march up
layer by layer applying the rule
\begin{equation}
    u_{i,j} =  \max \left\{ (u_{i-1,j-1} + \sleft),
                              (u_{i+1,j-1} + \sright)
                 \right\}
\end{equation}
This configuration can be thought of as the packing that would 
result from an infinitesimal gravitational force acting on the system.
(Strictly speaking, we set $u_{i,0}=0$, march upward, 
determine $U_0$ from the height reached, and translate 
all displacements accordingly.)

\begin{figure}[htbp]
  \centerline{ \epsfxsize=2.5in \epsfbox[0 0 182 182]{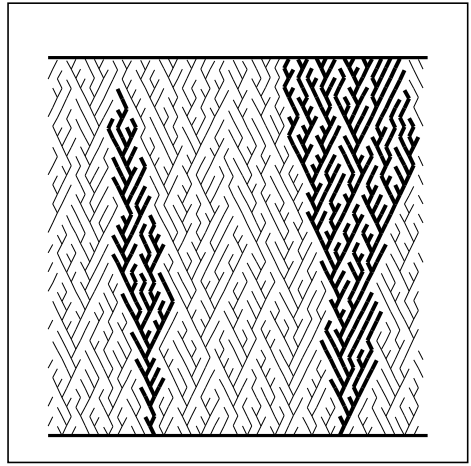} }
  \vspace{0.2in} \caption{A typical initial configuration.  All
  springs shown are at exactly zero compression.  Springs with
  equilibrium lengths too short to connect their pair of sites are not
  drawn at all.  The heavier lines are guides to the eye, highlighting
  two of the trees in this configuration.}  \label{fig:trees}
\end{figure}

A typical initial configuration is shown in Fig.~\ref{fig:trees}.
Sites joined by line segments are separated precisely by the
equilibrium length of the spring between them.  Springs that are not
long enough to connect their two sites are not drawn.  Already one can
see that there is a rich geometric structure hidden in this model,
quite similar to that described by Roux et al
\cite{roux87c}.  
The heavy lines in the figure highlight structures we call ``trees'', 
consisting of all sites that are connected to the same site 
on the first layer.  

We refer to the distance between the upper end of an {\em inactive}
spring and the site to which it will eventually connect as a ``gap''.
Note that sizes of the gaps reflect the structure of our
initialization algorithm.  We have moved all the grains to their
lowest possible position, concentrating all of the small gaps that
might be present in a chain into a single large gap at the top of a branch
of a tree.

To generate the force curve, we maintain a list of active sites and
their active bonds, and we keep track of changes in the tree structure
of the uncompressed chains as the floor is raised.  
For any fixed set
of active bonds, the $u_{i,j}$'s increase linearly with $\Delta$,
since all of the active springs are linear.  It is therefore a
straightforward linear problem to solve for the rate, $du/d\Delta$, at
which each $u_{i,j}$ advances.  Advancing $\Delta$ through one stage
involves solving for $u_{i,j}$ for some arbitrary increase in $\Delta$, 
for which we employ an iterative biconjugate gradient routine \cite{recipes},
using the current configuration of $u_{i,j}$'s as an initial guess for
the solution.  The solution is used to determine the rates of advance
of all the active sites.  The rate of advance of each inactive site is
equal to that of the active site that supports it through an
uncompressed chain.  The rates are used to determine the value of
$\Delta$ at which the network topology changes.  The $u_{i,j}$'s are
then updated according to the calculated rates, the list of active
sites or the tree structure is updated, and the process is repeated.

The updating of the network topology is necessitated either by the
closure of a gap or by the breaking of a bond.  The closure of a gap
can generate one of two types of events.  First, it can cause an
additional set of sites to be activated, which increases the size of
our active site list (and also increases the size of the linear
problem that must be solved on the next iteration).  Second, it can
result in a ``push'' event, in which an inactive branch of one tree is
simply transferred to another tree without becoming active.  Pushes
dominate the behavior of the system in the early stages of
compression, becoming more and more rare as compression continues and
the number of inactive sites decreases.

The breaking of a bond,
as mentioned in Section~\ref{subsec:features}, is the final
possibility for changing the network of active bonds.  
Breaks are generally rare events.  During full
compression of a $40\by 40$ system, which undergoes approximately 2000
events (pushes, chain additions, breaks), there
are typically about 5 breaks.  

The algorithm reaches completion when all bonds have been activated.
As explained above, further compression would be homogeneous 
with all $\fboth$'s increasing at precisely the same rate.
To compress one configuration on a 40x40 lattice, generating one
complete force curve, requires approximately one hour of computation
on a typical 200MHz workstation.  
We have accumulated data for various lattices with at most a
few thousand sites.  
Future efforts to optimize the algorithm should permit 
investigation of substantially larger lattices.

Despite our reliance on linear methods for evolving the system between
gap events, it must be emphasized that the absence of tensile forces
introduces a strong nonlinearity for larger increments in $\Delta$.
An alternative approach to following the entire evolution of the system
would be to solve directly for the configuration of the system at an arbitrarily
chosen value of $\Delta$ by minimizing the nonlinear
energy function of Eq.~(\ref{eq:Eij}).   
Such an approach might speed up the calculation even with linear springs,
and it would be essential if the springs were nonlinear
(e.g. Hertzian) under compression.  

\newpage
\section{The macroscopic force as a function of displacement}
\subsection{General remarks and a ``mean-field'' prediction}
Fig~\ref{fig:forcecurves} displays numerically computed curves of
$F$ vs. $\Delta$
for a few rectangular systems of various sizes.  
At a cursory level, the force curves appear to have the form 
$F(\Delta) \sim \Delta^{\nu}$, 
which correspond to lines of slope $\nu$ on our log-log plots.
Interestingly, this exponent depends strongly on system geometry.

Closer inspection of Fig.~\ref{fig:forcecurves}
reveals both expected and unexpected behavior.  
For any particular realization of the disorder,
there is a region, sometimes substantial, on the log-log plot 
for small values of $\Delta$ in which the force curve appears nearly linear
because it is dominated by a single chain.
There is also a crossover to linear behavior for large $\Delta$
since, as the fraction of active bonds approaches unity, 
the system approaches the linear limit
described in previous sections.
The intermediate regime is the one of interest,
and the behavior there is rather complex.
In many runs an exponent $\nu>1$ in the intermediate regime
can be readily identified.  
However, in other runs (see, e.g., Fig.~\ref{fig:forcecurves}c),
small-system statistics tend to obscure the phenomena.
It appears that completely reliable measurements
of the exponents 
in the various regimes will require bigger systems, 
beyond the reach of our current numerical codes.  
Nevertheless, we believe that
the results for systems of a few thousand grains
support the conclusions drawn below.
(See Ref~\cite{roux87a} for a discussion of a closely related model.)

In studying force curves for bead packs, several authors
have proposed a mean-field argument suggesting that for various
granular systems one should observe $\nu=\alpha+1$, where
$\alpha$ is the exponent of the single-contact force law.
Thus for our model the mean-field theory predicts $\nu=2$.
(For the reader's convenience, we summarize a version of the theory
in the Appendix.
The treatment there is similar to that of 
References~\cite{goddard90} and~\cite{roux87a}.)
The argument
is based on the assumption that the rates at which pairs of nearest
neighbor sites approach each other may all be taken to be
equal to the average rate of compression.

In Subsection C below we present numerical results for our model
showing that $\nu$ can be significantly less than 2
for some system geometries
and in Subsection D we seek to explain the failure of the
mean-field argument. 
In our view, this failure casts doubt on the argument's
applicability to real granular systems.
Useful supplementary information is covered in Subsections~B and~E.

\subsection{Analytic results for limiting cases}
It is instructive to consider two cases for which the force
curves can be explicitly calculated.

\begin{figure}[htbp]
  \centerline{ \epsfxsize=5.5in \epsfbox[0 0 521 234]{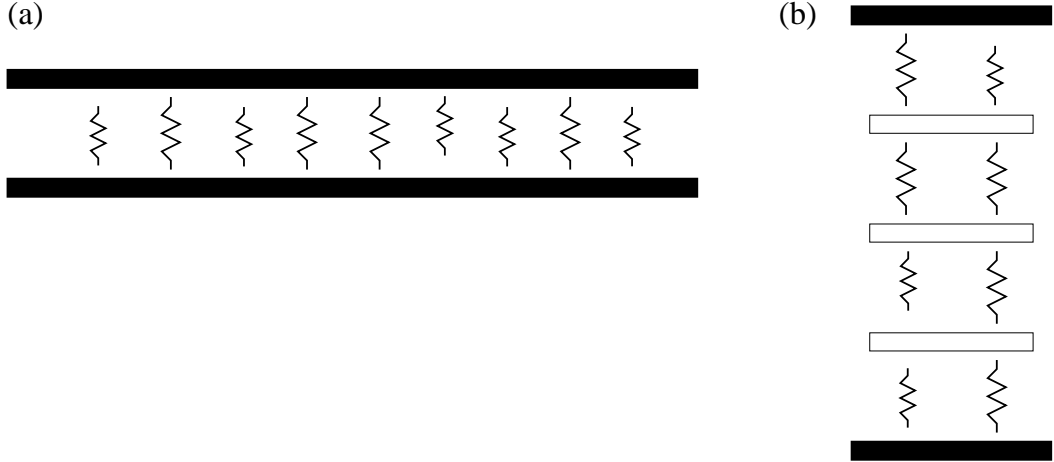} } 
  \vspace{0.2in} \caption{Limiting models for which 
  $F(\Delta)$ is easily computed analytically:
  (a) an infinite row of springs with random lengths ($N_y=1$);
  (b) an infinite column of single grains with two springs in every row ($N_x=1$).}
  \label{fig:limitmodels}
\end{figure}

{\em Case 1:} $N_y = 1$.  This system consists of only a single layer of 
random-length springs, as shown in Fig.~\ref{fig:limitmodels}a. 
Assuming the equilibrium spring lengths to be
uniformly distributed in a finite interval, one immediately
obtains $F\propto\Delta^2$.
In this case the mean-field argument (Appendix A) is exact,
as all gaps do close at the same rate.

{\em Case 2:} $N_x=1$.  Because of the periodic boundary conditions,
this case is equivalent to a single column of grains 
without periodic boundary conditions, as shown in Fig.~\ref{fig:limitmodels}b.
Again we take the distribution of equilibrium lengths to be
a square distribution of width $2 \epsilon$. 
In this case it is more convenient to fix the force $F$ and
compute the displacement $\Delta(F)$ since the compressive stress
on each grain must be the same.
Let us decompose the displacement 
\begin{equation} \label{eq:Dd}
 \Delta = \sum_{n=1}^{N_y} \delta_n ,
\end{equation}
where $\delta_n$ is the displacement of the $n$th spring.
When $F=0$, we have $\delta_n=0$ for each $n$,
the longer spring at each level being just at the threshold
of compression.
The growth of $\delta_n$ with $F$ depends only on $a_n$,
the difference between the equilibrium lengths of the two
springs in the $n$th layer. Specifically,
\begin{equation}
  \label{eq:dddF}
  \frac{d\delta_n}{dF} = \left\{ \begin{array}{ll}
  \vspace{1ex}
  \frac{1}{k}  & \mbox{if $0<F<ka_n$} \\
  \frac{1}{2k} & \mbox{if $ka_n<F$.}\end{array}\right.
\end{equation}
Note that the $a_n$'s are independent and all have the probability density
$P(a) = (2\epsilon-a)/2\epsilon^2$ for $0<a<2\epsilon$.

The expected value of $d\Delta/dF$ is defined as
\begin{equation}
  \label{eq:averagedddF}
  \left\langle\frac{d\Delta}{dF}\right\rangle = \sum_{n=1}^{N_y}\int_{0}^{2\epsilon} 
      \frac{d\delta_n}{dF} P(a_n)\; da_n.
\end{equation}
Using the fact that all the terms in the summation are equal
and performing the integration, we obtain $\langle d\Delta/dF\rangle$,
which in turn may be integrated to yield
\begin{equation}
  \label{eq:averagedelta}
  \left\langle \delta\right\rangle = \left\{\begin{array}{ll}
  f - \frac{1}{4}f^2 + \frac{1}{24}f^3  & \mbox{if $0<f<2$} \\
  \frac{1}{2}f + \frac{1}{3}            & \mbox{if $2<f$,}\end{array}\right.
\end{equation}
where $\delta\equiv\Delta /N_{y}\epsilon$ and $f\equiv F/k\epsilon$
are the nondimensionalized displacement and force per layer, respectively.
Eq.~(\ref{eq:averagedelta}) shows that, in this case,
there is no simple power-law behavior in the large $N_y$ limit.
Note that Eq.~(\ref{eq:averagedelta}) is approximately linear near $f=0$
(i.e., $\lim_{f\rightarrow 0^+} \left[d(\ln\delta)/d(\ln f)\right] = 1$),
and it is linear near and beyond $f=2$.
If we attempt to identify a single power law for intermediate values
of $f$, the natural choice is the derivative
$d(\ln \delta)/d(\ln f)$, 
evaluated at the point where this quantity is most slowly varying;
i.e., where 
$d^2(\ln \delta)/d(\ln f)^2$ vanishes.
This method yields $f\sim \delta^{\nu}$ with $\nu = 1.27...$,
where we have returned to displacement as the independent variable.

The analysis of Case 2 shows that it is possible for
the force curve to exhibit a region corresponding to a power less than 2.
It also shows that the emergence of a true power law 
should not be taken for granted in these systems.
We find, however, that in sufficiently wide systems, 
a power law does arise (cf. discussion below).

\subsection{Numerical results}
\begin{figure}[htbp]
  \centerline{ \epsfxsize=5.5in \epsfbox[44 52 503 706]{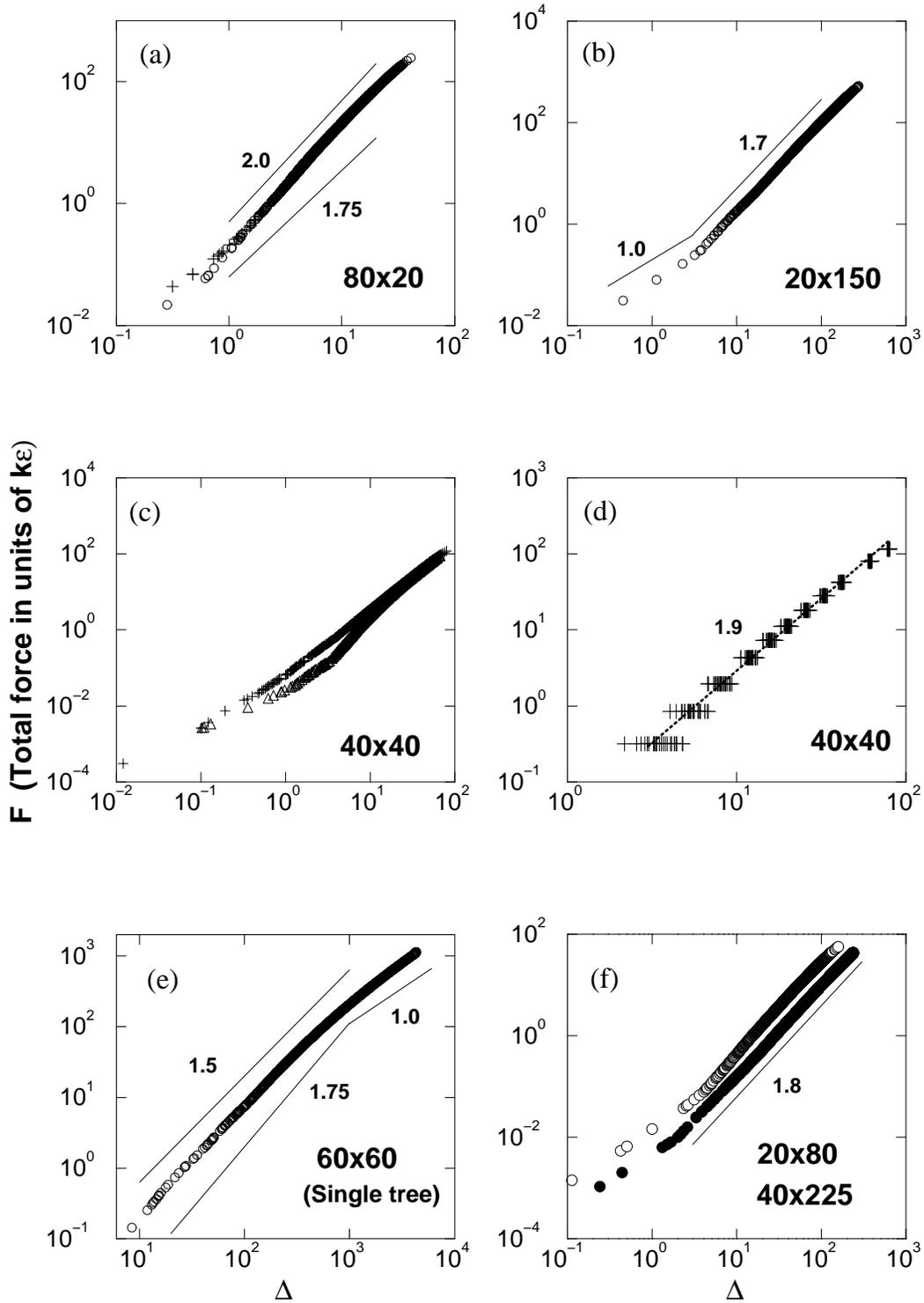} } 
  \vspace{0.2in} \caption{Plots of
  $F(\Delta)$ for several different system sizes.  Line segments with
  labels indicating their slopes are added as guides to the eye.
  Labels in the lower right corners indicate system size.  
  (a) A short, wide system showing an exponent near 2.  
  (b) A tall, narrow system showing an exponent near 1.7.  
  (c) A square system showing apparently different behavior 
  for different realizations of the quenched random equilibrium lengths.  
  (d) Data from 25 runs of a $40\by 40$ system.  For clarity,
  points are plotted only at 10 discrete values of $F$.
  (Note the different horizontal scale from (c)).
  The dotted line through the data is a power-law fit to the points shown,
  excluding the two highest forces.  The exponent is 1.9.
  (e) A single tree showing an exponent of 1.5.
  (f) Two different system sizes with equal values of $N_x N_y^{-0.66}$,
  both showing an exponent of 1.8.
  The open circles represent one completed run of a $20\by 80$ system.
  The filled circles represent one partially completed run of a $40\by 225$ system.}
  \label{fig:forcecurves}
\end{figure}

{\em Short, wide systems:} Fig.\ref{fig:forcecurves}a shows two
typical force curves for an $80\by 20$ system.  It appears here that
there is a regime in which $\nu\agt 2$, followed by the expected
crossover to $\nu=1$ at large $\Delta$.  This observation lends some
support to the mean field argument and is consistent with the claim of
Gilabert et al. \cite{gilabert87}, who studied the electrical analogue
of our model.  (See Section~\ref{sec:electrical} below.)

{\em Tall, narrow systems:}
In a $20\by 150$ system, the exponent $\nu$ appears to shift noticeably.
The best fit to the curve shown in Fig.~\ref{fig:forcecurves}b is
$\nu=1.7$ in the intermediate regime of interest.  
The error in this measurement is estimated to be $\pm 0.1$ on the 
basis of fits made with different choices for which points to exclude
from the intermediate regime.
The data clearly rule out $\nu=2$.  

{\em Roughly square system:}
An estimate of $\nu=1.9$ is obtained from data on $40\by 40$ systems.   
Fig.~\ref{fig:forcecurves}c shows data from two $40\by 40$ systems.
The apparent lack of consistent behavior
is presumably due to finite size effects.
Nevertheless, using data from 25 runs, an estimate of $\nu=1.9$ 
can be obtained as shown in Fig.~\ref{fig:forcecurves}d.
To obtain the dotted line in the figure, the data in the intermediate
regime were fit to a power-law;  data from the initial linear regime
and data from the two largest forces in the figure (where we expect a
crossover to linear growth) were excluded in making this estimate for
the exponent $\nu$.

\vspace{0.1in}
We suggest that the variation of exponent with aspect ratio can
be traced to the tree structure in the system.
Specifically, we conjecture that if most of the force is transmitted
within a single tree, then a smaller exponent will be observed,
whereas if the force is spread over many trees, a larger exponent.
We interpret the following numerical experiment as support for this conjecture.
Starting with a $60\by 60$ grid,
we removed all sites lying outside the $90^\circ$ cone emanating
from the center of the bottom boundary.
We then measured $F(\Delta)$ for this reduced system.
In this geometry all the active sites at every stage of compression
are connected to the floor at the same point, so all the
active bonds are contained within a single tree.
As shown in Fig.~\ref{fig:forcecurves}e the exponent $\nu$ 
is close to $1.5$, indicating
that the behavior within a single tree has a different character
than in a system with many trees.

To explore this issue further, we measured the rate at
which trees expand as a function of height {\em in the initial configuration}.  
(Roux et al. have investigated a similar, but not identical, model 
\cite{roux87c}.)  
While building the initial configuration, we
keep track of the root of the tree associated to each site, 
thereby counting the number $T(z)$ of trees that survive up to
layer $z$. 
Since only initial configurations are involved ---
not their subsequent evolution ---
rather large systems can be simulated.
For a system of width 50,000 with 
data collected for up to $10^7$ layers,
we obtained an excellent power law fit $T(z) \approx A z^{-\gamma}$, 
with $\gamma = 0.66$ and $A=6.5\times 10^4$, 
with only a slight deviation for very small heights ($z < 10$)
and very large heights (corresponding to $T(z)<5$).
Note that $A=1.3 N_x$.  
(Incidentally, since $T(1)=N_x$, it is clear that the power law fit
must be inaccurate for $z$ of order $1$.)
Generalizing from this system to one of arbitrary width,
we estimate that in the initial configuration of an
$N_x\by N_y$ system, approximately $1.3 N_x N_y^{-\gamma}$ trees
will reach the top layer.

Let us speculate on possible consequences of this estimate.
The rates of advance of all sites within a single tree are correlated,
since compression of the bottom bond of the tree affects the rates
of all the branches above it.
We therefore expect the sequence of gap closings to depend upon 
the initial tree structure.
Assuming that the system obeys a simple anisotropic scaling law,
we conjecture that in large systems the exponent $\nu$
is a function of $N_x N_y^{-\gamma}$ alone.
Limited support for this conjecture comes from two additional
runs on $20\by 80$ and $40\by 225$ systems.
In both cases, $N_x N_y^{-0.66}\approx 1.1$,
and in both cases we observe power-law behavior with $\nu\approx 1.8$,
as shown in Fig.~\ref{fig:forcecurves}f.
A more rigorous test
would require runs on significantly larger systems,
for which a more efficient code is needed.
Note that the $40\by 225$ system was already too large for us
to run to completion; the curve in Fig.~\ref{fig:forcecurves}f ends
well before the crossover to the linear regime occurs.
Note also that in the $40\by 225$ system, a clean power-law regime
extends over at least two decades in $\Delta$ and three decades in $F$,
giving us some confidence that a true power-law regime does exist 
in large systems.

A consequence of our scaling conjecture is that increasing the system
size while keeping the aspect ratio {\em fixed} should result in
exponents that eventually approach that of a wide, short system.  As
we have shown that $\nu=2$ in the $\infty\by 1$ limit (Case 1,
Section~IVB), we expect to observe the mean-field result $\nu=2$ in
very large, approximately square systems as well.

\subsection{Why the mean field argument fails}
\begin{figure}[htbp]
  \centerline{ \epsfxsize=5.5in \epsfbox[45 462 502 669]{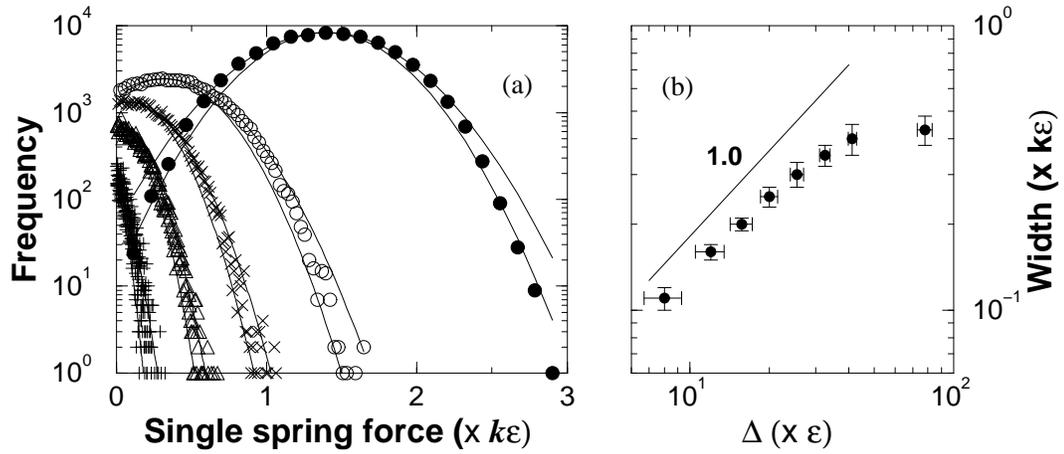}} 
  \vspace{0.2in} \caption{Distribution of spacings between nearest
  neighbor grains for several stages of compression of a $40\by 40$ system.
  (a) From narrowest to broadest, the curves represent 
  active bond densities of 0.2, 0.4, 0.6, 0.8 and 1.0, which
  correspond to ensemble averaged total forces of 
  0.85, 4.3, 11.2, 27.9, and 116.1 $\times k\epsilon$. 
  The data represents averages of 25 runs.
  (b) The width of the fitted Gaussians as a function
  of $\Delta$ on a log-log scale.
  Horizontal error bars indicate the range of values of $\Delta$ obtained
  for a given density of active bonds in different runs.
  The thin lines in the left plot are the Gaussians used for the
  maximal and minimal width estimates at each stage.
  The line of slope 1.0 on the right is a guide to the eye.}  \label{fig:Pofx}
\end{figure}

The key assumption in the mean-field argument is
that nearest neighbor sites approach each other at the mean compression rate.
In terms of the probability density for inter-site distances,
the assumption is expressed quantitatively
in Eq.~(\ref{eq:Pxdelta}) of the appendix.
which asserts that
the shape and width of the distribution of inter-site distances
are independent of the compression $\Delta = N_y\delta$.
As discussed in the appendix, in our model the probability density
$P(x,\delta)$ is well defined only for $x>0$;
i.e., for bonds whose springs are compressed.
However, even restricting our attention to the active bonds,
we find that our data are inconsistent with the above assumption.
Specifically, in Fig.~\ref{fig:Pofx}a we show the probability densities
for the (nonzero) forces $f$, 
where $f = \max\{kx,0\}$, 
at various stages of compression of a $40\by 40$ system,
and it is clear that the distribution broadens as $\delta$ increases.
More quantitatively, in Fig.~\ref{fig:Pofx}b we plot the widths
of the best fits to the data by Gaussian distributions restricted to $\{x>0\}$,
for several values of $\delta$.

The broadening of $P(x,\delta)$, which occurs during 
the early and intermediate regime of compression,
is a simple consequence of force balance at
branching points of chains of active bonds.
Consider a site at which
three active bonds meet, forming  a ``Y''
(either right-side-up or upside-down). 
Because there are no tensile forces,
the force in the unpaired branch
of the Y is greater than either of those in the paired branches.
Moreover, as the Y is further compressed, the rate at which
the force in the unpaired branch increases must equal
the {\em sum} of the rates of increase of the forces in the paired branches.
Since the larger force evidently increases more rapidly
than the two smaller forces (the forces almost always being
increasing functions of $\Delta$), the distribution broadens.

During the late regime of compression, the addition
of active bonds will convert Y's to X's, for which
any correlation between force and bond compression rate 
must be more subtle.
Indeed, when all of the bonds are activated, 
all bonds compress at exactly the same rate
and the force distribution simply shifts uniformly to
higher forces -- the mean field assumption becomes exact.
Thus we expect to observe a broadening of the force distribution
during early and intermediate stages of compression,
with a rate that decreases as the density of Y's becomes smaller.
This is precisely the behavior displayed in Fig.~\ref{fig:Pofx},
where the width of the positive $x$ portion of the distribution 
is seen to grow roughly
as a power of $\Delta$ less than unity for small $\Delta$
and level off at high compression.
A quantitative calculation of the rate of broadening would
require a detailed understanding of the statistics of
branching in the active network and is beyond the scope of this work.

Let us argue that
the broadening of the force distribution generically leads
to an exponent $\nu$ smaller than the mean field value.
Suppose, for definiteness, that $P(x,\delta)$
has a sharp leading edge at $x_1(\delta)$;
i.e., that $P(x,\delta) = 0$ for $x > x_1(\delta)$
and $\lim_{x\rightarrow x_{1}^{-}} P(x,\delta)$ is
bounded away from zero.
The existence of Y's in the active bond network leads
to an advance of the leading edge $x_1(\delta)$ that is
more rapid than the advance of $\delta$ itself.
Taking into account the formation of new contacts
and branch points during compression, and
assuming simple asymptotic behavior, we expect that
$x_1(\delta) \sim \delta^{\beta}$ with $\beta < 1$. \cite{betanote}
Thus, to lowest order in $\delta$, we find
$F = \int_{0}^{x1} P(x,\delta) kx\;dx \sim \delta^{2\beta}$,
and therefore $\nu\le 2$.
Corrections to this exponent should become significant
for $\delta$ of order unity, which is also the order of
the width of $P(x)$.
This expectation is consistent with Fig.~\ref{fig:forcecurves},
where it can be seen that the crossover to the linear regime for large
$\Delta$ occurs for $\Delta\gtrsim N_{y}/5$, which is 
the last half decade in the plots.

These qualitative results
do not depend on the linearity (under compression) 
of the springs in our model, 
nor do they depend on the dimensionality of the model.
They should also apply to the {\em vertical} forces in a system
with horizontal degrees of freedom, 
providing only that the creation of new branch points is sufficiently common.
Thus the mean-field argument seems problematic for physical bead packs.
The analysis shows that  
new contact formation can lead
to exponents smaller than the mean field value,
an important point in light of
other mechanisms proposed to explain
experimental observations of this exponent. \cite{goddard90,degennes96}

\newpage
\subsection{Relation to random resistor networks}\label{sec:electrical}
The analogy between the elastic properties of a network of linear
springs and the electrical properties of the same network of resistors
has been exploited in numerous studies \cite{guyon90}.
In order to clarify the relation between our model and others,
and particularly with random networks near the directed percolation threshold,
let us consider the analogy in some detail.
We will see that there are good reasons to be skeptical of the applicability
of percolation model results to our model, but there is also an intriguing
numerical coincidence.

There is a formal identity between our equations for mechanical equilibrium 
\begin{eqnarray}\label{eq:electric1}
 \fboth & = & -k(u_{i,j} - u_{i\pm 1,j} - \sboth)
\end{eqnarray}
and the electrical equations
\begin{eqnarray}\label{eq:electric2}
 \Iboth & = & \frac{1}{R}(V_{i,j} - V_{i\pm 1,j} - \vboth). 
\end{eqnarray}
In the latter equation, $\Iboth$ specifies the current between sites
that are joined by a resistance $R$ in series with a
battery generating a potential difference $\vboth$.
Note that the resistors' conductance
plays the role of the springs' stiffness. 
The property that the springs function only under compression 
can be modeled in the electrical system by the insertion of perfect diodes, 
all directed ``downward'', in series with each resistor.  
Roux et al. studied this very model \cite{gilabert87}, 
but did not report results for varying aspect ratios or single trees 
and did not study its relation to the \qmodel, 
which had not yet been introduced.

A system with batteries is substantially more complicated than a simple, 
randomly diluted resistor-diode network.
To relate the randomly diluted network at the directed percolation threshold
to our model just beyond the initial linear regime,
one must first assume that the network of active springs in our model has 
the same structure as the current carrying paths in a directed network 
of resistors placed at random on the lattice.  
One must also assume a relation between $\Delta$ of our model
and the probability $p$ that a bond exists in the percolation system.
The natural assumption here would be that the probability 
$p-p_c$ is proportional to $\Delta$, where $p_c$ is the critical value
for percolation, since $\Delta=0$ is the point where a single force chain 
(or current carrying path) first forms and on average the bonds in our model
are compressed by an amount proportional to $\Delta$ \cite{guyon90}.
In our system, however, the batteries play two roles.
First, they generate potentials that affect the current distribution
even when all the diodes are forward biased,
which would correspond to a trivial, homogeneous state of the simple resistor network.
Perhaps more importantly, however, they determine {\em which} diodes
will be forward biased for a given applied potential difference
across the whole network.
This dynamical process of selecting the current carrying paths may or may not
yield structures well-modeled by the random addition of resistive links
in a percolation model.

In spite of the difficulties in establishing a connection, 
it is interesting to compare our results 
for the power law obeyed by the stiffness $dF/d\Delta$ 
to the conductivity exponent 
obtained from the theory of directed percolation in
randomly diluted resistor networks \cite{redner82,arora83}.
The conductivity exponent for directed percolation has been
calculated both numerically and using renormalization group methods
and appears to be approximately $0.7\pm0.1$,
\cite{redner82,arora83}
which corresponds to a value of $1.7\pm0.1$ for the exponent $\nu$.
The fact that this agrees with our measurements on tall, 
narrow systems deserves further study.
Other authors have investigated additional details of the 
statistics near threshold in our model and argued that the
system is closely related to the percolation one. \cite{roux87b}

\section{Statistics of forces on individual grains}
We now consider the statistics of forces transmitted by individual springs,
exploring, in particular, the relation of our results
to the \qmodel\ of Coppersmith et al.\ \cite{qmodel1}.
Our model is simpler than the granular packings
the \qmodel\  was intended to describe.
However, our model
would appear to be a better candidate for description
by the \qmodel\  than the original granular packings,
 since we have removed the tensorial stresses 
from our system but still have contacts whose
formation is governed by quenched randomness.  

In the \qmodel\ in two dimensions, 
$q_{i,j}$ refers to the fraction of the force on
site $(i,j)$ from above that is transmitted to its
neighbor at $(i-1,j-1)$, the complementary fraction $1-q_{i,j}$
being transfered to site $(i+1,j-1)$.
One assumes that 
each $q_{i,j}$ is a random variable that 
(1) is independent of the other $q$'s and 
(2) has the same distribution $\eta(q)$ at every site, independent
of the force supported by that site.

\begin{figure}[htbp]
  \centerline{ \epsfxsize=5.5in \epsfbox[57 245 444 668]{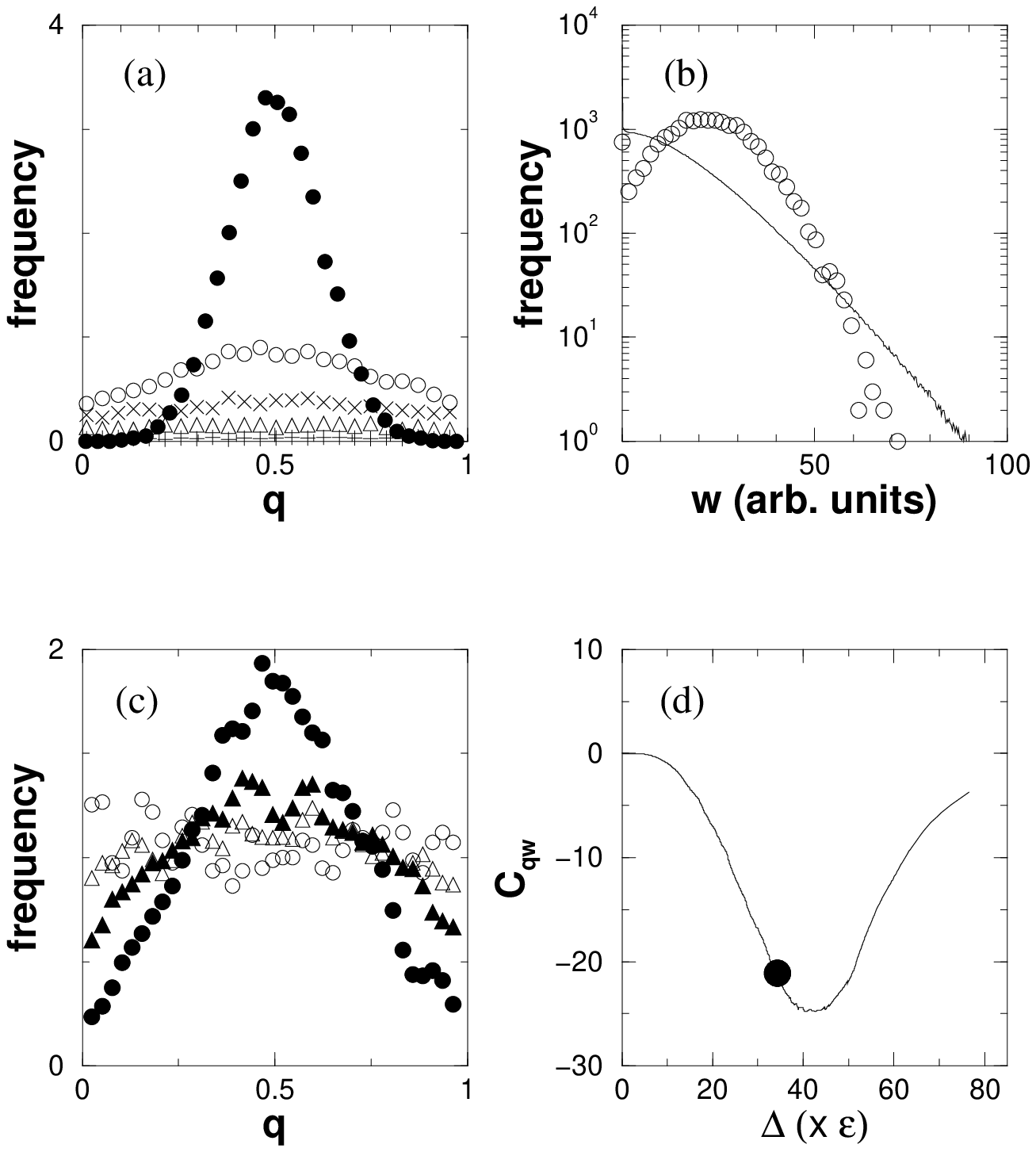}} 
  \vspace{0.2in} \caption{Distribution of $q$ values and comparison
  with predictions of the $q$-model.  (a) The frequency distribution
  of $q$'s averaged over 25 configurations of a $40\by 40$ system.  
  The different symbols indicate different stages of compression,
  with larger values at $q=1/2$ corresponding to larger compressions.
  All distributions are normalized to unity, but 
  the points at $q=0$ and $q=1$ are off scale in all cases except for
  the fully compressed one. 
  (b) Comparison of the distributions of total force supported by a single
  grain in our model and in the $q$-model with a similar frequency
  distribution of $q$ values.  (See text for details.)  
  The open circles are averages over the same 25 configurations used
  to generate the open circles in (a). 
  The heavy dots are averages over 100 configurations of the \qmodel.
  (c) Separation of the $q$ distribution from one curve in (a) into components
  corresponding to different levels of supported force.  The data in
  (c) correspond to the open circles in (a).  
  The bins used are 
  $(0,W)$ (open circles),
  $(W,2W)$ (open triangles),
  $(2W,3W)$ (filled triangles), and
  $(3W,\infty)$ (filled circles),
  where $W$ is the average force supported by a single grain.
  (d) The correlation function
  $C_{qw}\equiv\langle(q_{ij}-\frac{1}{2})^2 (w_{ij}-\langle
   w_{ij}\rangle)\rangle$.
  The data shown are from one realization of a $40\by 40$ system.  
  The heavy dot indicates the point corresponding to the data in (b) and (c).}
  \label{fig:qdist}
\end{figure}

Analytical studies of the \qmodel\ show that
the distribution of forces supported
by a single grain has an exponential tail at large forces
whenever $\eta(q)$ is non-vanishing at $q=1$. \cite{qmodel1,claudin98}
By contrast,
the force distributions in our model, 
plotted in Fig.~\ref{fig:Pofx},
show no evidence of exponential tails.
At no stage during the compression does it appear that the \qmodel\
distributions are a good match for the distributions we observe.
In particular, consider the fourth curve from the
left in Fig.~\ref{fig:Pofx}, which is made at a compression
for which almost all {\em sites} are active but there remains an
appreciable fraction of inactive {\em bonds}.
The distribution of $q_{i,j}$'s directly measured 
from our data in this regime 
(see Fig.~\ref{fig:qdist}a) is reasonably represented by
\begin{equation} \label{eq:eta}
\eta(q)=0.1\left[\delta(q)+\delta(1-q)\right]+0.8-0.2\cos(2\pi q),
\end{equation}
for which 20\% of the $q$'s are either 0 or 1.
The \qmodel\  would predict an exponential tail in the 
single grain force distribution. \cite{qmodel1}
Even for the small systems we study here, the exponential tail would
be clearly distinguishable from the rapid decay we observe.
Fig.~\ref{fig:qdist}b compares numerical results from our model and
from a simulation of the \qmodel\  with $\eta(q)$ given by Eq.~(\ref{eq:eta}).
For the \qmodel\  simulation, 
forces on the top row were chosen randomly from a 
uniform distribution on the interval $(0,2)$.  

To understand the discrepancy between our results and the predictions
of the \qmodel, we examine our data as regards assumption (2) above:
i.e., that the distribution of $q$ at each site is independent of the
force supported by that site.
Fig.~\ref{fig:qdist}c shows the distribution of $q$'s obtained from
equilibrium configurations of our model corresponding to the same
conditions as in Fig.~\ref{fig:qdist}a, but separated according to the
force supported by the site.  
Different symbols in the plot indicate
different levels of force as described in the figure caption.  
It is clear that the contribution to $\eta(q)$ from larger forces 
is peaked more strongly about 1/2 and has little weight near $0$ or $1$, 
which explains why the \qmodel\  prediction fails for this system.  
As a quantitative measure of the correlation, we have computed
the covariance
$C_{qw}\equiv\langle(q_{ij}-\frac{1}{2})^2 (w_{ij}-\langle
w_{ij}\rangle)\rangle$ 
for our model, 
where $w_{ij}\equiv \fleft+\fright$ 
and the averages are performed over space 
for one realization of a $40\by 40$ system.  
(The square in this definition is
necessary because left-right symmetry guarantees that a correlation
function linear in $q-1/2$ would vanish.)  
$C_{qw}$ vanishes identically in the \qmodel\ 
since $q_{i,j}$ and $w_{i,j}$ are independent in that model.  
As shown in Fig.~\ref{fig:qdist}d, $C_{qw}$ develops a
significant negative value when the system is under compression,
indicating that larger
forces are associated with $q$'s closer to 1/2.  
The heavy dot in the figure indicates the point 
corresponding to the data in parts (b) and (c).

\section{Conclusions}
We briefly summarize the results of our study of the scalar model,
discuss its generalization to three dimensions,
and finally draw two conclusions concerning the implications
for real systems or more realistic models.

In the context of a toy model, we have tested arguments that have
been applied to stresses in static, non-cohesive granular materials.
Further study of the model is needed,
especially simulations of larger systems,
but already two important facts have been established.
(1) Correlations in the stress configuration are responsible
for substantial effects, both at the level of single-grain forces
and that of macroscopic stresses.
(2) Depending on the (properly scaled) aspect ratio,
new contact formation may play a decisive role in determining
the macroscopic stress-strain relationship,
with sufficiently tall systems showing power-law behavior
with a nontrivial exponent.
Regarding (1), we have identified two important effects:
(i) the distribution of forces broadens under compression
because of force balance constraints at sites
where three active bonds meet; and
(ii) larger forces tend to divide more evenly between supporting grains
as a result of the dynamical process that determines the 
structure of the active bond network.
Theories that neglect these correlations fail when applied
to our model.

One may wonder whether qualitatively new features might appear
in a 3D generalization of our model in which 
the vertical direction is taken to be 
the 111 direction of a simple cubic lattice.
As a preliminary check,
we have measured the statistics of trees in the initial configuration 
and observed their general morphology.  
The number of surviving trees decays as $z^{-1.2}$
and the trees remain relatively compact.
In other words, the diameter of surviving trees grows as $z^{0.60}$
(compared to $z^{0.66}$ in two dimensions) and 
the branches of separate trees do not become heavily entangled 
with each other.  
On this basis, we conjecture that the essential physics 
of the stress-strain relation will not be
qualitatively different in three dimensions.  In particular,
we expect a variation in the exponent $\nu$ with the scaled
aspect ratio.

We draw two general conclusions.
First, it appears that the mechanism for the broadening of the vertical force
distribution should be present in the full tensorial problem as well,
and therefore should be re-examined as a possible explanation for 
the observation of smaller exponents than those derived from the mean-field argument.
Experiments measuring the dependence of the exponent $\nu$ on
aspect ratio would be especially interesting.
Second, the success of the \qmodel\ in predicting the distribution
of single-grain forces in bead packs stands in need of explanation.
The \qmodel\  has already been criticized for its neglect of
tensorial stresses \cite{socolar98}. 
One may have expected it to provide a better description
of the stresses in a scalar model of the type we have studied,
but this is not borne out by our results.

We thank Sue Coppersmith and Long Nguyen for helpful conversations,
and also Scott Zoldi for his suggestions.
This work was supported by NSF through grants DMS-98-9803305 (DGS and MGS)
and DMR-94-12416 (JESS).

\newpage
\appendix
\section{The mean-field argument predicting $\nu=2$}
In this appendix, we summarize a mean-field argument that was
developed to describe forces during the compression of a
bead pack between parallel plates (See, e.g.,
Refs.~\cite{goddard90,roux87a}.)

We define the \textit{overlap} between two adjacent beads as 
\mbox{$(d_0-d)$} where $d$ is the distance between their centers and
$d_0$ is the nominal distance at which the beads touch but exert no
force.  Note that because of the minus sign, the overlap is positive
for beads in contact, and negative for beads not in contact.

We introduce the random variable, $X$ to be the overlap between two
randomly chosen adjacent beads.  Thus the
sample space for $X$ includes both the choice of configuration (spring
lengths in our model), and the choice of a pair of adjacent beads. Of course $X$
also depends on the displacement $\Delta$ of the floor plate. (As in
Section I, we normalize so that nonzero forces start at $\Delta=0$).
We rescale the independent variable, defining $\delta= \Delta/N_y$
where the bead pack is $N_y$ layers thick.  Let $P(x,\delta)$ be the
probability distribution for $X$.  The average force per spring,
$\sigma$, is given by;
\begin{equation}
  \label{eq:sigma}
  \sigma (\delta) = \int_{-\infty}^{\infty} f(x) P(x,\delta)\; dx,
\end{equation}
where $f(x) =  \max\{kx,0\}$.  Since
\begin{equation}
   \label{eq:fxo}
   f(x)=0\quad {\rm for}\quad x<0,
\end{equation}
the integral may be restricted to $(0,\infty)$.  Of course, by our
normalizations,
\begin{equation}
  \label{eq:Pxo}
  P(x,0) = 0\quad {\rm for}\quad x>0,
\end{equation}
so $\sigma(0)=0$.  The rescaled variable $\delta=\Delta/N_y$ equals
the average change in the overlap $X$ resulting from motion of the
floor; thus,
\begin{equation}
  \label{eq:dmd0}
  \int_{-\infty}^{\infty} x P(x,\delta)\; dx = \delta+\delta_0,
\end{equation}
where $\delta_0$ is the average overlap when $\delta=0$.  

The first main assumption of the mean-field argument is a considerable
strengthening of (\ref{eq:dmd0}): One assumes that the shape and width
of the probability distribution are independent of $\delta$, i.e.,
\begin{equation}\label{eq:Pxdelta}
        P(x,\delta) = P(x-\delta,0).
\end{equation}
Substituting (\ref{eq:Pxdelta}) into (\ref{eq:sigma}) and rescaling
(\ref{eq:fxo},\ref{eq:Pxo}), we deduce that
\begin{equation}
  \sigma = k \int_{-\delta}^{0}(x+\delta)P(x,0)dx,
\end{equation} 
which is equivalent to Eq.~(1.2) of Ref.~\cite{goddard90}.  

The second main assumption of the mean-field argument is that
$P$ is continuous in both arguments and
\begin{equation}
  \label{eq:pofo}
  \lim_{x\rightarrow 0^-} P(x,0) = C,
\end{equation}
where $C$ is a finite positive constant.
Combining the two assumptions, we find that
\begin{equation}
   \sigma = k \int_{-\delta}^{0} (x + \delta) \left[C+{\cal O}(x)\right] dx,
\end{equation}
which immediately yields $\nu=2$.  
Under assumption (\ref{eq:Pxdelta}), the only way to
accommodate a value of $\nu$ less than 2 is to posit that
the limit in Eq.~(\ref{eq:pofo}) is infinite.

Note that in our model, the
distribution for $X$ is not uniquely defined for $X < 0$ since
inactive grains are free to relocate slightly.  Despite this
ambiguity, in section IVD we are able to test hypothesis
(\ref{eq:Pxdelta}) by focusing on positive values of $X$.

\newpage
\begin{center}
{\Large Figures}
\end{center}

\begin{thebibliography}{10}
\bibitem{betanote}
If the active bond network were fixed, we would obtain
$\beta = 1$, but the advance
of the {\em trailing} edge of the distribution
would invalidate the analysis;
for a fixed network of active bonds, 
the system is linear and $\nu=1$.
\end{thebibliography}
\end{document}